\documentstyle{article}

\begin{document}

\title{The influence of the Wolff's cluster structure 
on the determination of the
critical dynamic exponent}
\author{P. R. A. Campos\thanks{
prac@ifqsc.sc.usp.br}, N. G. F. de Medeiros\thanks{
ngfm@ifqsc.sc.usp.br} and R. N. Onody\thanks{
onody@ifqsc.sc.usp.br} \\
{\small {\em Departamento de F\'{\i}sica e Inform\'{a}tica } }\\
{\small {\em Instituto de F\'{\i}sica de S\~{a}o Carlos} }\\
{\small {\em Universidade de S\~{a}o Paulo - Caixa Postal 369} }\\
{\small {\em 13560-970 - S\~{a}o Carlos, S\~{a}o Paulo, Brasil.}}}
\date{}
\maketitle

\baselineskip=16pt

\begin{abstract}
Recently we reported some interesting features of the Wolff's algorithm
behavior
 when applied to the site-bond-correlated Ising model.
Our main
 results were that a stronger correlation diminishes the 
autocorrelation time but it does not change or affect
the critical dynamic exponent.
In this paper, we analyse the Wolff's cluster structure and how it varies
with the spatial correlation. The fractal dimensions
are determined for several values of the magnetic atoms concentration 
and spacial correlation, giving an estimative of the critical dynamic exponent.

\vspace{3cm}


\end{abstract}

\thispagestyle{empty} 

{\normalsize \newpage }

{\normalsize With the advent of the cluster algorithms, a great improvement
in the simulations of magnetic spin systems has been possible (Swendsen
1987, Wolff 1989). When applied to the
usual dilute Ising model (Hennecke and Heyken 1993), such algorithms exhibit
a performance which is better than those of single-spin flip techniques.
It was verified that as the concentration of non-magnetic atoms increases
the corresponding autocorrelation time decreases.
Recently, we obtained very similar results about the
performance of Wolff algorithm in the site-bond correlated Ising
model (Campos and Onody 1997, Campos and Onody 1998). It was shown that the
algorithm becomes even more robust in the presence of spatial correlation.
This a very important feature, since for the single-spin flip algorithms
(Metropolis, Glauber) the correlation weakens their performance, and so a
poor statistical data can result. In this letter we deal with the analysis
of the Wolff's clusters structures and their importance in obtaining the
critical dynamic exponent. }

{\normalsize The site-bond correlated Ising model (hereafter, SBC model) was
proposed in a way to explain additional effects not predicted by the simple
dilute Ising model, and experimentally verified in the antiferromagnetic
compound {\it KNi}$_{x}${\it Mg%
}$_{1-x}${\it F}$_{3}${\it \ } (Aguiar, Engelsberg and
Guggenheim 1986). The main attribute of this model is the existence of a local
spatial correlation. In the SBC model, the presence of 
impurities in the neighbourhood of a given (nearest neighbors)
pair of magnetic atoms can {\it modify
 the strength} 
of the exchange coupling constant between these two atoms. Moreover, in
the limit of maximum correlation }$\alpha ${\normalsize , the exchange 
interaction can
 be even suppressed. }

{\normalsize The SBC model Hamiltonian is defined as follows (Aguiar,
Moreira and Engelsberg 1986) }

{\normalsize \smallskip }

{\normalsize {\it 
\begin{equation}
H=-\sum_{i,\delta }J_{i,i+\delta }(\sigma _{i}\sigma _{i+\delta }-1)
\end{equation}
}where }$\sigma _{i}=\pm 1${\normalsize \ and }$\delta ${\normalsize \
denotes an elementary lattice vector. The exchange interaction }$%
J_{i,i+\delta }${\normalsize \ is given by }

{\normalsize \smallskip }

{\normalsize {\it 
\begin{equation}
J_{i,i+\delta }=J\varepsilon _{i}\varepsilon _{i+\delta }[(1-\alpha
)\varepsilon _{i-\delta }\varepsilon _{i+2\delta }+\alpha ]
\end{equation}
}}

{\normalsize \noindent where }$J>0.${\normalsize \ The random variables }$%
\varepsilon _{i}${\normalsize \ can take the following values: one with
probability }$C${\normalsize \ and zero with probability }$1-C${\normalsize %
, where }$C${\normalsize \ is the concentration of magnetic atoms. The
parameter }$\alpha ${\normalsize \ correlates the interaction between sites }%
$i${\normalsize \ and }$i+\delta ${\normalsize \ with the magnetic occupancy
of the sites }$i-\delta ${\normalsize \ and }$i+2\delta .${\normalsize \ The
uncorrelated dilute Ising model is re-obtained in the limit }$\alpha =1.$%
{\normalsize \ For }$0<\alpha <1,${\normalsize \ the bond between }$i$%
{\normalsize \ and }$i+\delta ${\normalsize \ is only weakened by the
absence of a magnetic atom at }$i-\delta ${\normalsize \ or }$i+2\delta $%
{\normalsize . The limit }$\alpha =0${\normalsize \ corresponds to the
maximum correlation, i.e., two magnetic first neighbor sites are connected
by an active bond only if their nearest-neighbor sites along the line
joining them are also present. This maximum limit (}$\alpha =0${\normalsize %
) defines at T=0 a new percolation problem, with the concentration threshold 
}$p_{c}=0.7405${\normalsize \ (Campos, Pessoa and Moreira 1997). For }$%
\alpha \allowbreak \neq 0${\normalsize \ the critical concentration is like
that the usual site percolation. }

{\normalsize Following the Fortuin-Kasteleyn (Fortuin and Kasteleyn 1969)
arguments and the definition of the model hamiltonian, we can derive the
bond activation probabilities required in the construction of the clusters.
The probabilities are defined as follows (Campos and Onody 1997): }

{\small \smallskip }{\normalsize 
\begin{equation}
\left\{ 
\begin{tabular}{l}
$p_{i.i+\delta }=0$ \ \ \ \ \ \ \ \ \ \ \ \ \ \ \ \ \ if \ $\sigma _{i}\neq
\sigma _{i+\delta }$ \\ 
$p_{i.i+\delta }=1-e^{-2K}$ \ \ \ \ \ \ \ if \ $\sigma _{i}=\sigma
_{i+\delta }$ \ and \ $\varepsilon _{i-\delta }\varepsilon _{i+2\delta }=1$
\\ 
$p_{i.i+\delta }=1-e^{-2\alpha K}$ \ \ \ \ \ if \ $\sigma _{i}=\sigma
_{i+\delta }$ \ and \ $\varepsilon _{i-\delta }\varepsilon _{i+2\delta }=0$%
\end{tabular}
\right.  \label{prob}
\end{equation}
\ between sites $i$ and $i+\delta .$ }

{\normalsize We simulated the SBC model for some values of concentration }$C$
{\normalsize and correlation }$\alpha $. {\normalsize Due to the large
geometric fluctuations, an accuracy statistical study and a great
computational efforts were performed. In Table 1, we show our results for
the critical dynamical exponents }$z${\normalsize \ for some values of the
parameters. These results were obtained by a log-log plot of the
autocorrelation time }$\tau ${\normalsize \ versus }$L${\normalsize , since
as expected }$\tau ${\normalsize \ obeys a power law like }$\tau \sim L^{z}$%
{\normalsize .}

\bigskip 

{\small The values of the critical dynamical exponent z for some values of
correlation }$\alpha ${\small \ and concentration C.}{\normalsize \ } 
\begin{equation}
\begin{tabular}{||l|l|l||}
\hline\hline
& $C=0.80$ & $C=0.70$ \\ \hline
$\alpha =0.1$ & $0.26$ & $0.10$ \\ \hline
$\alpha =0.3$ & $0.34$ & $0.11$ \\ \hline
$\alpha =0.5$ & $0.38$ & $0.33$ \\ \hline
$\alpha =1.0$ & $0.40$ & $0.38$ \\ \hline\hline
\end{tabular}
\end{equation}

{\normalsize As we can see, there is an increasing of the }$z${\normalsize \
exponent with augmented values of correlation. This behavior is verified for
all concentration values. In order\ to compare the performance of the Wolff
dynamics with others, we need to rescale }$\tau $ {\normalsize (Wolff 1989),
since one Wolff step has a computational cost of the cluster size }$\left|
S\right| ${\normalsize . So, we define the Wolff autocorrelation time by the
relation}

\begin{equation}
\tau _{w}=\tau \frac{\left\langle \left| S\right| \right\rangle }{CL^{d}}%
\end{equation}
{\normalsize where }$d${\normalsize \ is the lattice dimensionality, and }$%
CL^{d}${\normalsize \ is the total magnetic mass of the system. Here, we
restrict our study to the case }$d=2${\normalsize . \ Like }$\tau $%
{\normalsize , the Wolff time }$\tau _{w}${\normalsize \ also exhibits a
power law of the form \ }$\tau _{w}\sim L^{z_{w}}${\normalsize , where }$%
z_{w}${\normalsize \ is the Wolff critical dynamical exponent.}

{\normalsize Another way to estimate the }$z_{w}${\normalsize \ values,
consists in regarding a power law behavior for the mean cluster size} $%
\left\langle \left| S\right| \right\rangle \sim L^{Df}${\normalsize , where }%
$Df$ {\normalsize is the fractal dimension of the Wolff clusters. In this
case, we can calculate }$z_{w}$ {\normalsize through the equation}

\begin{equation}
z_{w}=z-(d-Df).
\end{equation}
{\normalsize In the case of the usual Ising model, the advantage of this
procedure is the removal of one source of error, since } $\left\langle
\left| S\right| \right\rangle =\left\langle \chi \right\rangle \sim L^{\frac{%
\gamma }{\nu }}${\normalsize , where }$\chi $ {\normalsize is the magnetic
susceptibility and }$\frac{\gamma }{\nu }${\normalsize \ is the ratio of the
exactly known Ising exponents (Wolff 1989).\qquad }

\bigskip

{\normalsize In a recent work, we verified no dependence of the }$z_{w}$%
{\normalsize \ values with the parameters }$C${\normalsize \ and }$\alpha $%
{\normalsize \ (Campos and Onody 1998). We found a critical dynamical
exponent value around zero. This result indicates that the Wolff algorithm
circumvents almost completely the critical slowing down phenomenon, and so
making possible a more accuracy statistical analysis, since there is no
statistical correlation between the sucessive configurations generated by
the algorithm. }

{\normalsize But this an intriguing thing. With this assumption, the
scenario presented for the }$z${\normalsize \ exponent is not valid for }$%
z_{w}${\normalsize . The only way to support these results regards the
changes occurred in the fractal dimension of the Wolff clusters with the
parameter }$\alpha ${\normalsize \ in order to keep constants the }$z_{w}$%
{\normalsize \ values . The study of the cluster structures are our main
focus on this work.}

\bigskip

{\normalsize In Figure 1, we present two pictures of the Wolff clusters
produced in different generations. On the top of the figure, we have }$%
\alpha =0.1${\normalsize \ and }$C=0.7${\normalsize . On the bottom, }$%
\alpha =1.0${\normalsize \ and the same value for }$C${\normalsize . The
configurations were generated in the respective critical temperature. As we
can see, the clusters for }$\alpha =0.1${\normalsize \ are more compact than
those for }$\alpha =1.0${\normalsize . This means that the value of }$(d-Df)$%
{\normalsize \ is smaller for }$\alpha =0.1${\normalsize \ than for }$\alpha
=1.0${\normalsize . For this reason, the tendency of increasing of }$z$%
{\normalsize \ with }$\alpha ${\normalsize , is not predicted for the }$z_{w}
${\normalsize \ values.}

\bigskip

{\normalsize We acknowledge Conselho Nacional de Desenvolvimento
Cient\'{i}fico e tecnol\'{o}gico (CNPq) and Funda\c{c}\~{a}o de Amparo a
pesquisa do estado de S\~{a}o Paulo (FAPESP) for financial support. }

{\normalsize \bigskip }

{\normalsize \newpage }

\end{document}